\documentclass{aa}

\usepackage{graphicx}
\usepackage{ulem}
\usepackage{color}
\usepackage{txfonts}
\usepackage{natbib}
\usepackage[dvipsnames]{xcolor}
\bibpunct{(}{)}{;}{a}{}{,}

\begin{document}

\title{Limiting magnetic field  for minimal deformation of a magnetised neutron star}

\author{R.O. Gomes\inst{1} \and H. Pais\inst{2,3} \and V. Dexheimer\inst{4} \and C. Provid\^encia\inst{2} \and S. Schramm\inst{1 \thanks{Deceased}}}

\institute{
Frankfurt Institute for Advanced Studies, Frankfurt am Main, Germany.  \and
CFisUC, University of Coimbra, P-3004-516 Coimbra, Portugal. \and
INESC TEC, Campus da FEUP, P-4200-465 Porto, Portugal. \and
Department of Physics, Kent State University, Kent OH 44242 USA. \\ {\tt rosana.gomes@ufrgs.br, hpais@uc.pt, vdexheim@kent.edu, cp@uc.pt} 
}
\offprints{H. Pais}
\date{Received xxx Accepted xxx}

\abstract{}
{In this work, we study the structure of neutron stars under the effect of a poloidal magnetic field and determine the  limiting largest  magnetic field strength that induces a deformation such that  the ratio between the polar and equatorial radii does not exceed $2\%$. We consider that, under these conditions, the  description of magnetic neutron stars in the spherical symmetry regime is still satisfactory.}
{We  describe different compositions of stars (nucleonic, hyperonic, and hybrid), using three state-of-the-art relativistic mean field models (NL3$\omega\rho$, MBF, and CMF, respectively) for the microscopic description of matter, all in agreement with standard experimental and observational data. The structure of stars is described by the general relativistic solution of both Einstein's field equations assuming spherical symmetry and Einstein-Maxwell's field equations assuming an axi-symmetric deformation.}
{We find a limiting  magnetic moment of the order of $2\times 10^{31}$Am$^2$, which corresponds to magnetic fields of the order of 10$^{16}$ G at the surface and $10^{17}$ G at the centre of the star, above which the deformation due to the magnetic field is above $2\%$, therefore, not negligible. We show that the intensity of the magnetic field developed in the star depends on the EoS, and, for a given baryonic mass  and fixed magnetic moment, larger fields are attained with softer EoS. We also show that the appearance of exotic degrees of freedom, such as hyperons or a quark core, is disfavored in the presence of a very strong magnetic field.  As a consequence, a highly magnetized nucleonic star may suffer an internal conversion due to the decay of the magnetic field,  which could be accompanied by a sudden cooling of the star or a gamma ray burst.}
{}


\titlerunning{Limiting $B-$field for minimal deformation}
\authorrunning{Gomes et al.}

\maketitle

\section{\label{intro} Introduction}

Neutron stars are one of the possible remnants of supernova explosions that are triggered by the gravitational collapse of intermediate mass stars. During the collapse, because of angular momentum and magnetic flux conservation, the rotation frequencies and magnetic fields of these stars are exceptionally amplified, reaching values of $P \sim 1$ s and $B_s \sim 10^{12}$ G, respectively. Also, because of the extreme densities reached inside these objects, neutron stars provide a unique environment for investigating fundamental questions in physics and astrophysics. 
 In particular, a class of objects named magnetars possess surface magnetic fields that are even more extreme than regular pulsars, of the order of $B_s \sim 10^{13} - 10^{15}\,\mathrm{G}$, appearing in nature in the form of Soft gamma repeaters (SGRs) and Anomalous X-ray pulsars (AXPs).  The nature of these objects was explained in the magnetar model proposed by Thompson $\&$ Duncan \citep{Thompson:1995gw,Thompson:1996pe}, which interprets them as neutron stars that present highly energetic gamma-ray (SGRs) and X-ray (AXPs) activity, powered by their strong magnetic fields decay. For a broad review on magnetars from theoretical and observational points of view, see \cite{Mereghetti:2015asa}, \cite{Kaspi:2017fwg}, and references therein. 
Although the internal magnetic fields of neutrons stars can not yet be accessed by observations, Virial theorem arguments \citep{Lai1991} predict that the internal magnetic fields can reach values up to $B_c \sim 10^{18}$ G in stellar centers, which agree with realistic general relativity calculations \citep{Bocquet:1995je,Cardall:2000bs,Frieben:2012dz,Chatterjee:2014qsa}

In the past, the effects of strong magnetic fields were studied, first, separately on the equation of state (EoS) of neutron star matter \citep{Broderick:2000pe,Broderick:2001qw} and in a fully general relativistic formalism on the structure of neutron stars \citep{Bonazzola:1993zz,Bocquet:1995je}. Only much later, the effects of strong magnetic fields were studied self-consistently on the EoS and structure of neutron stars \citep{Chatterjee:2014qsa,Franzon:2015sya}. The latter concluded that magnetic field effects on the EoS of baryons and quarks do not play a significant role for determining the macroscopic properties of stars with central values $B_c \lesssim 10^{18}$ G (as expected from simple order of magnitude estimates), although the same cannot be said about direct magnetic field effects on the macroscopic stellar structure. In particular, it was shown that neglecting deformation effects by solving general relativity spherically symmetric solutions (TOV) \citep{Tolman:1939jz,Oppenheimer:1939ne}, leads to an overestimation of the mass and an underestimation of the equatorial radius of stars \citep{Gomes:2017zkc}. This happens because, in this case, the extra magnetic energy that would deform the star (when the proper formalism is applied) is being added to the mass due to the imposed spherical symmetry.

The particle population of the core of stars is also directly affected by the presence of strong magnetic fields. The main reasons for this being the shift of the particle onset density to higher densities if magnetic field effects are taken into account in the EoS \citep{Chakrabarty:1996te,Chakrabarty:1997ef,Broderick:2000pe,Broderick:2001qw,Yuan1999,Suh:2000ni,Lai1991} and, also, the decrease of stellar central densities  \citep{Franzon:2015sya,Franzon:2016urz,Gomes:2017zkc,Chatterjee:2018ytb}. Depending on the intensity of the magnetic fields considered, exotic particles such as hyperons, delta resonances, meson condensates and quarks can vanish completely from their core, changing, for example, the neutrino emission of these objects \citep{Rabhi2009a}. In other words, magnetic field decay over time can lead to a repopulation of stars, similarly to rotational spin down \citep{Negreiros:2011ak,Bejger:2016emu}, playing an important role on key questions such as the hyperon puzzle  {\citep{Zdunik:2012dj,Chatterjee2015}}, the Delta puzzle \citep{Cai:2015hya,Drago2016}, hadron-quark phase transitions \citep{Avancini2012,Ferreira2013,Costa2013,Roark:2018uls,Lugones18} and stellar cooling \citep{Sinha:2015bva,raduta2017,Negreiros18,Fortin18a,Patino:2018tyq}.

In the light of the aforementioned results from studies of magnetic neutron stars modeled in a self-consistent formalism, it is clear that a careful determination of the magnetic field threshold beyond which a spherical symmetry for stars is no longer valid is needed, together with the determination of the lowest magnetic field that makes exotic particles vanish from the core of stars. We address these two questions in this work, first, by identifying the intensity of the magnetic fields relevant for modifying the macroscopic properties of neutron stars. 
For doing so, we compare the structure of magnetic neutrons stars using a formalism that allows stars to deform due the effect of a poloidal magnetic field and test different values of the magnetic dipole moment of the star, including the zero dipole case. Second, we investigate the conditions that give rise to the conversion of a hadronic star to a hyperonic or hybrid one due to the decay of the magnetic field. We use three different stellar compositions (nucleonic, hyperonic, and hybrid) and different relativistic mean field models, in order to make our results more general. 

 In the present work, we  consider a poloidal magnetic field configuration  for non-rotating stars.
 The  joint effect of a toroidal magnetic field and rotation on the structure of neutron stars was carried out in \cite{Frieben:2012dz}. It was shown that neither purely poloidal nor purely  toroidal  magnetic  field configurations are stable and that stability requires   twisted-torus  solutions \citep{Ciolfi2014}. In \cite{Uryu2014}, the authors have numerically obtained stationary solutions of relativistic rotating stars considering strong mixed poloidal and toroidal magnetic fields. In particular, the presence of a dominant toroidal component is essential, for instance, to describe an increase of the inclination angle of a NS \citep{Lander2018}.  
Therefore, a more complete description of magnetised neutron stars should be addressed in the future.

In order to consistently describe magnetars in a general relativistic framework, we use the formalism implemented in the publicly available Langage
 Objet pour la RElativit\'e Num\'eriquE (LORENE) library \citep{Lorene2,Bonazzola:1993zz,Bocquet:1995je,Bonazzola1998,gourgoulhon2012} in its present online version. It solves the coupled Einstein-Maxwell field equations in order to determine stable and stationary magnetised configurations of stars by assuming a poloidal magnetic field distribution.
The metric  used for the polar-spherical symmetry is the Maximal-Slicing-Quasi-Isotropic (MSQI) \citep{gourgoulhon2012,Franzon:2015sya}, which allows the stars to deform by making the metric potentials depend on the radial $r$ and angular $\theta$ coordinates with respect to the magnetic axis. As this formalism does not account for the hydrodynamical generation of electromagnetic fields, these are introduced via macroscopic currents, which are free parameters in the calculation. Alternatively, we can determine the currents by fixing the stellar magnetic dipole moment, which is the conserved quantity in our system.
 
In \cite{Bocquet:1995je}, the authors present  the first numerical results of the coupled Einstein-Maxwell equations for highly magnetised rotating neutron stars. They study the structure of magnetized stars considering for the core several neutron matter  \citep{Diaz1985,Haensel1981,Pandharipande1971}, one polytropic, and a hyperonic matter \citep{Bethe1974} EoS. Non unified crust-core EoS are used, with BPS for the outer crust and BBP or NV (see Sec. 4.1 of \cite{Salgado1994a}) for the inner crust.
In our study, we analyse again the effect of the magnetic field on magnetized non-rotating stars using LORENE and  considering three state-of-the-art EoS, NL3$\omega\rho$ (hadronic), MBF (hyperonic), and CMF (hybrid). As in \cite{Bocquet:1995je}, for the outer crust, we take the BPS EoS, but the inner crust EoS is built from a Thomas-Fermi pasta calculation using the NL3$\omega\rho$ model, so that the  inner crust-core EoS is unified for this EoS.

It was shown in  \cite{Bocquet:1995je} that the deformation of a star is only significant for $B>10^{14}$ G, that the maximum stellar mass increases with the magnetic field, and that the maximum allowed magnetic field is of the order of $10^{18}$ G. In the present work, we more thoroughly quantify the deformation created by the magnetic field and  find that the limiting magnetic field that causes a relative deformation of $2\%$ is of the order of $10^{16}$ G at the surface and $10^{17}$ G at the center. We also show that the largest effects  created when increasing the magnetic field are seen on the stellar radius and not on the mass, which is in agreement with the results of \cite{Chatterjee:2014qsa}. A different conclusion was drawn in \cite{Bocquet:1995je} because  the maximum magnetic dipole moment considered was above 10$^{32}$ Am$^2$, generating strong effects also on the stellar maximum mass.  For magnetic dipole moments below $10^{32}$ Am$^2$ (our case), the mass is not significantly affected when compared to the zero-field case. We do not include magnetic field effects in the EOS, as previous studies concluded that magnetic fields do not play significant role on  the core EoS, (\cite{Chatterjee:2014qsa}), though more recently, it has been shown that they do have a non-negligible role  in  the outer crust EoS \citep{Kondratiev2001,Potekhin:2012pq,Chamel:2012uz,Chamel:2015fna,Stein:2016tad,Potekhin:2017ufy} and in the inner crust EoS  \citep{fang16,fang17,Fang:2017zyz}. 
  In what follows, in section II, we present  the EoS models used in the study. 
Our results for different families of stars and for a single star are shown in section III. Finally, in section IV, some conclusions are drawn.

\section{\label{Methods} Equations of State}

 The full EoS used in this work are constructed with an outer crust, an inner crust, and a core. The outer crust is merged with the inner crust at the neutron drip line, and the inner crust is matched with the core EoS at the density for which the pasta geometries melt, i.e., the so-called crust-core transition.
For the outer crust, we use the Baym-Pethick-Sutherland (BPS) EoS \citep{bps}, and for the inner crust, we perform a Thomas-Fermi pasta calculation \citep{grill14,pais16} using the relativistic mean field NL3$\omega\rho$ model \citep{pais16}.

 We point out that more up-to-date outer crust EoS have been calculated, however,  it has been shown in  \cite{Fortin16} that the use of the BPS EoS for the outer crust or more recent EoS, such as the ones discussed in \cite{Ruester:2005fm}, will practically not affect the mass and radius of neutron stars. Also the authors of \cite{Sharma2015} have shown that the behavior of BPS EoS is very similar to the one of  BCPM \citep{Sharma2015}, or the one of BSk21 \citep{Pearson2012,Potekhin2013a}. Since BPS is a well known and frequently used EoS,  we choose to consider it in the following calculations. 

The EoS of the magnetized outer crust has been calculated in \cite{Potekhin:2012pq}, and applied in \cite{Potekhin:2017ufy} to study the cooling of magnetized neutron stars, or in \cite{Chamel:2012uz,Chamel:2015fna}, where the authors have demonstrated that the magnetic field could affect the mass of the outer crust.  However,  no EoS for the magnetized inner crust is currently publicly available.  In a consistent calculation, we should have considered an unified EoS of magnetized nuclear matter but we believe the error we introduce by not using the magnetized outer crust EoS is within the uncertainties of not using a completely unified EoS.

In order to make our results as general as possible, we use three  different models for the core EoS, each with a different stellar composition: nucleonic, hyperonic, and hybrid.  For the nucleonic core EoS, we consider the  NL3$\omega\rho$ model, composed of $npe$ homogeneous matter \citep{pais16}. This model fulfills constraints coming from microscopic neutron matter calculations, such as chiral effective models \citep{hebeler13}, and also generates two-solar-mass neutron stars.  For the hyperonic core EoS, we use the many-body force (MBF) \citep{Gomes:2014aka,Dexheimer:2018dhb} model with $npe\Lambda$ matter; and, for the hybrid one, we take the chiral mean field (CMF) model with $npe\mu \Lambda\epsilon q$ matter, taking into account chiral symmetry restoration and allowing for the existence of a mixed phase  \citep{Dexheimer:2009hi}. 
One should note that all of the {EoS} used in this work are calculated without taking into account microscopic magnetic field effects, since it has been shown in previous works \citep{Chatterjee:2015pua,Franzon:2015sya,Dexheimer:2016yqu,Gomes:2017zkc} that the magnetic field does not significantly affect the EoS or stellar central magnetic fields $B_c \lesssim 10^{18}$ G.  However, note that in some more recent studies,  it was shown that strong magnetic fields of the order of magnitude observed in stars can have important  effects in  the outer crust EoS and its properties \citep{Kondratiev2001,Potekhin:2012pq,Chamel:2012uz,Chamel:2015fna,Stein:2016tad,Potekhin:2017ufy} or in the inner crust EoS  \citep{fang16,fang17,Fang:2017zyz}. The nuclear matter properties at saturation density of the models discussed in this work are displayed in Table \ref{Hmodels}. Their astrophysical properties are shown in the following section.

\begin{table}
  \caption{\label{Hmodels} Nuclear properties for symmetric matter at saturation density for the three EoS models used in this work. The columns are saturation density $\rho_{0}$, binding energy per nucleon, $B/A$,  effective mass of the nucleon, $M^*$, incompressibility modulus, $K_0$, symmetry energy, $E_{sym}$, and the slope of the symmetry energy, $L$. All quantities are given in MeV, except for the saturation density, which is given in fm$^{-3}$. }

\begin{center}
\begin{tabular}{ccccccc}
 \hline
 \hline
Model & $\rho_{0}$ & $B/A$ &  $M^*$ &  $K_0$  & $E_{sym}$ & $L$  \\
  \hline 
 NL3$\omega\rho$ & $0.148$ & $-16.24$ & 559 & $270$  & $31.7$ & $55$  \tabularnewline
 MBF-$\omega \varrho$ & $0.149$ & $-15.75$ & $620 $ & $297$ & $26.4$  & 46   \tabularnewline
CMF         & 0.150 & $-16.00$ &  629  & 297  & 29.6 & 88  \tabularnewline 
  \hline\hline
  \end{tabular}
\end{center}
\end{table}


\section{\label{Bfield} Results }

In this section we begin by calculating sequences of magnetic stars with different central densities and baryonic number, using the LORENE library for fixed values of the magnetic dipole moment, $\mu$. This quantity is related to the radial component of the magnetic field, $B_r$, by: 
$$\frac{2\mu \cos\theta}{r^3}=B_r\mid_{r\rightarrow \infty} \, ,$$ 
which provides us with different magnetic field strength distributions in each stellar sequence.  
In a second moment, we fix the stellar baryonic mass and only change the value of the stellar magnetic dipole moment.

Our main goal in this work is to determine the maximum value of the magnetic field intensity for which neutron stars  can still be described by spherical equilibrium solutions of Einstein's equations, i.e the TOV equations, in a reasonable good approximation. 
The stellar matter EoS is described within the three models presented in the previous section and the effect of the magnetic field on several properties of magnetised stars is discussed. 

\subsection{Families of stars}

\begin{figure}[!htbp]
 \begin{tabular}{c}
\includegraphics[width=0.45\textwidth]{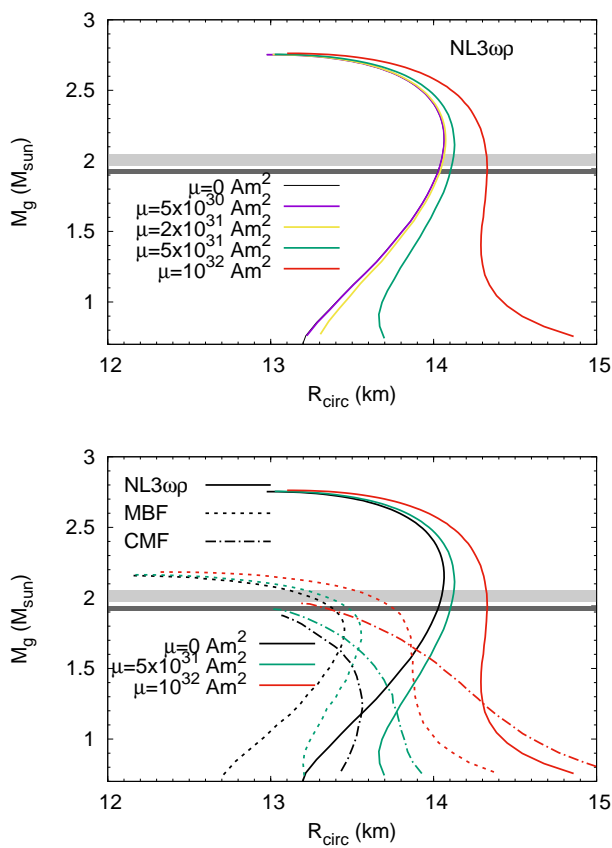} 
 \end{tabular}
  \caption{The mass-radius relation for the NL3$\omega\rho$ model for several values of the magnetic dipole moment, $\mu$,  (top), and for the three models considered in this paper  with three different values of the magnetic dipole moment (bottom). } 
\label{fig1}
\end{figure}

\begin{figure}[!htbp]
 \begin{tabular}{c}
\includegraphics[width=0.45\textwidth]{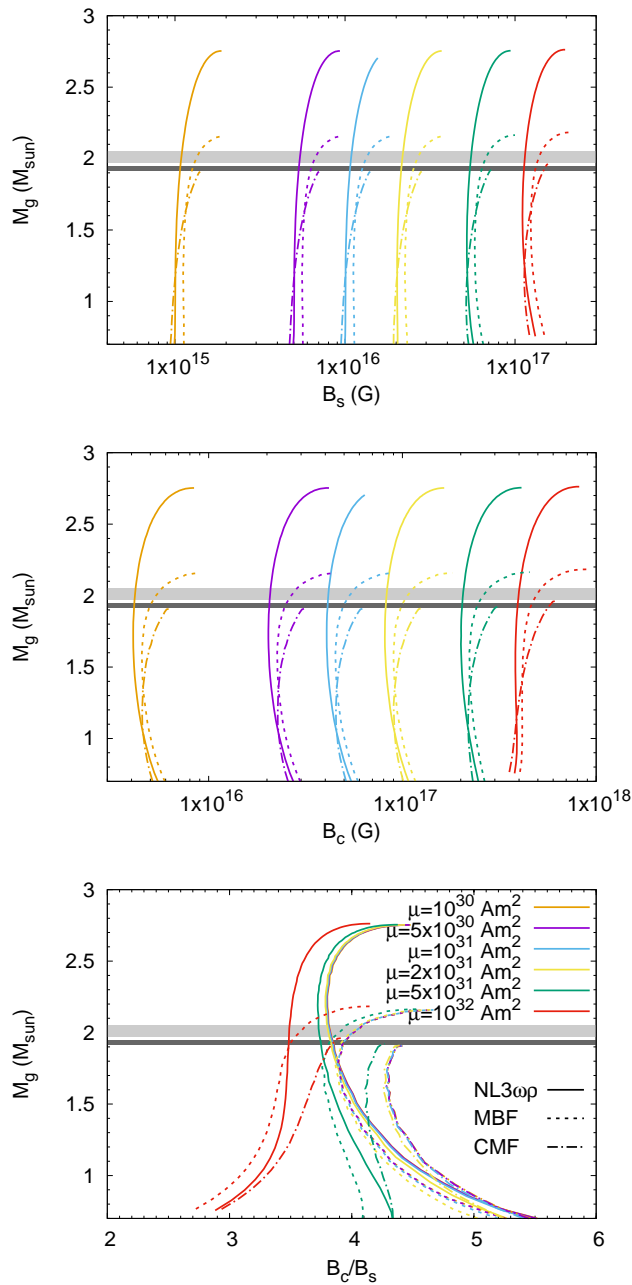} 
 \end{tabular}
  \caption{The gravitational mass of several families of stars with different values of the magnetic dipole moment, $\mu$, as a function of the surface polar magnetic field at the pole $B_s$ (top), the central magnetic field $B_c$ (middle), and the ratio between the central and surface magnetic fields, $B_c/B_s$, (bottom), for the NL3$\omega\rho$ (solid), MBF (dashed) and CMF (dot-dashed) models.} \label{fig2}
\end{figure}

\begin{figure}[!htbp]
 \begin{tabular}{c}
\includegraphics[width=0.45\textwidth]{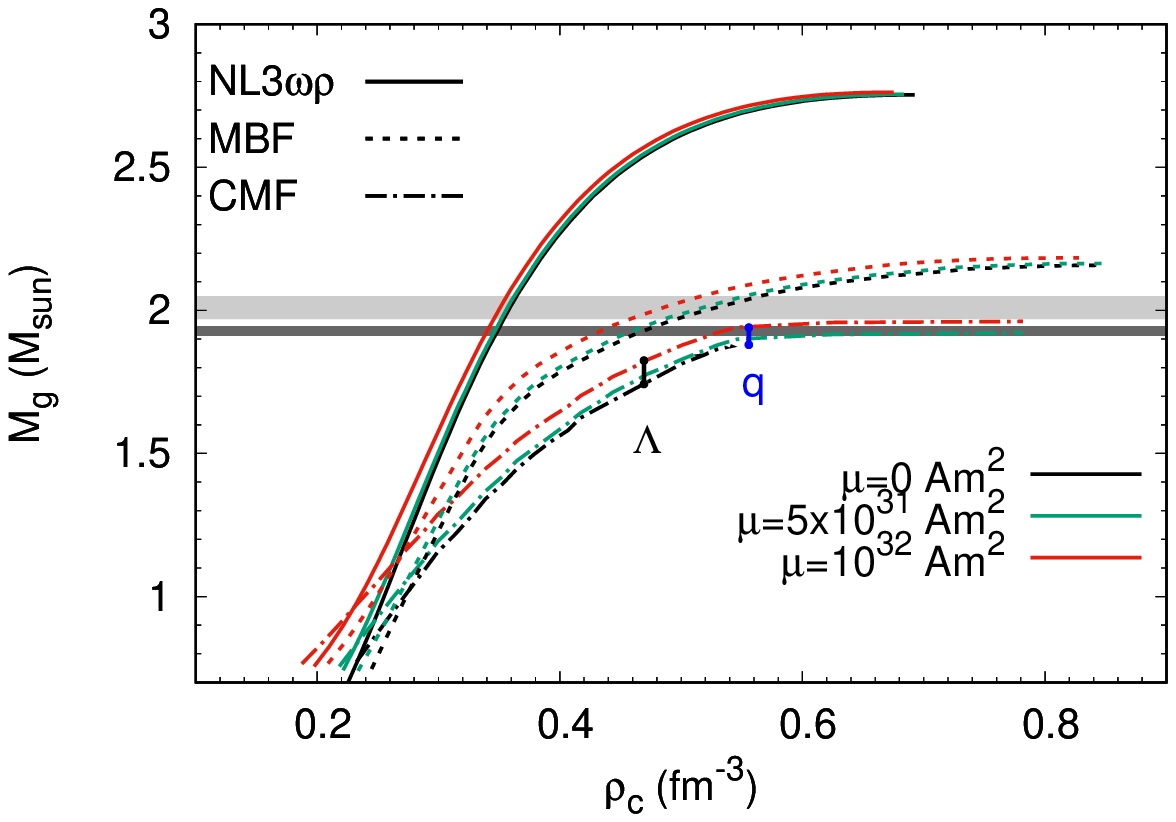} 
 \end{tabular}
  \caption{The gravitational mass of several families of stars with different values of the magnetic dipole moment, $\mu$, as a function of the central density for the three  models considered in this work. The onset densities of $\Lambda$ and quarks is also shown for the CMF model.}
 \label{fig3}
\end{figure}

\begin{figure}[!htbp]
 \begin{tabular}{c}
\includegraphics[width=0.45\textwidth]{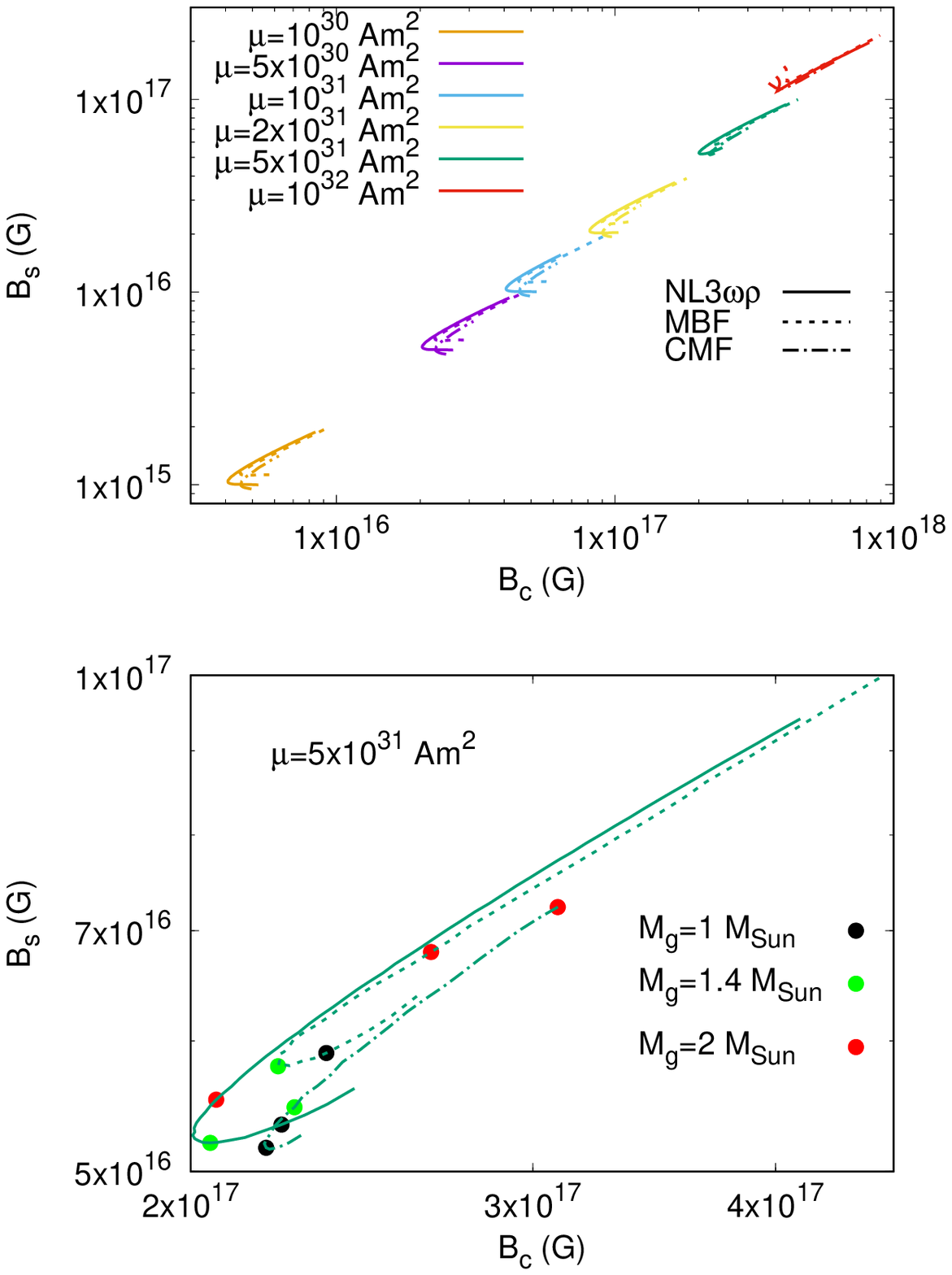} 
 \end{tabular}
  \caption{(Top) The correspondence between the central and surface (at the pole) magnetic fields, $B_c$ and $B_s$, respectively, for the three models. (Bottom) The same as the above panel but only for $\mu=5\times 10^{31}$ Am$^2$. The circles show three  values of the gravitational mass: $M=1$ M$_\odot$ (black), $M=1.4$ M$_\odot$ (green), and $M=2$ M$_\odot$ (red). The red circle for the CMF model represents the star with the maximum mass, $M=1.92$ M$_\odot$. }
\label{fig4}
\end{figure}

\begin{figure}[!htbp]
 \begin{tabular}{c}
\includegraphics[width=0.45\textwidth]{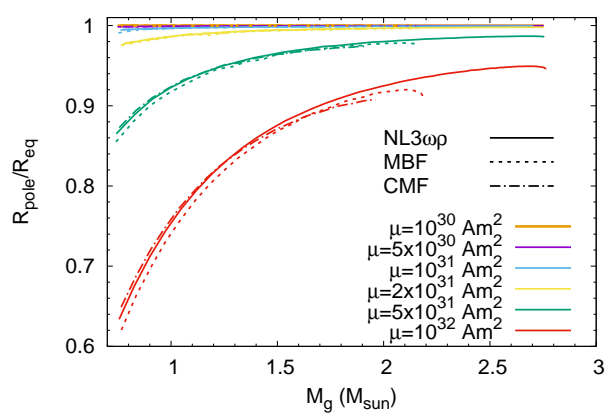} 
 \end{tabular}
  \caption{Deviation from spherical symmetry, given by the ratio between the stellar radius at the pole and at the equator, as a function of the gravitational mass for the three  models.   } 
\label{fig5}
\end{figure}

\begin{figure}[!htbp]
 \begin{tabular}{c}
\includegraphics[width=0.45\textwidth]{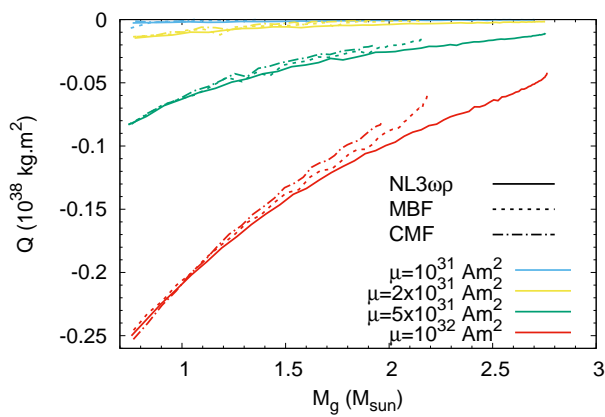} 
 \end{tabular}
  \caption{Quadrupole moment of stars as a function of the gravitational mass for the three  models.}    
\label{fig6}
\end{figure}

\begin{figure*}[thbp]
 \begin{tabular}{c}
\includegraphics[width=0.99\textwidth]{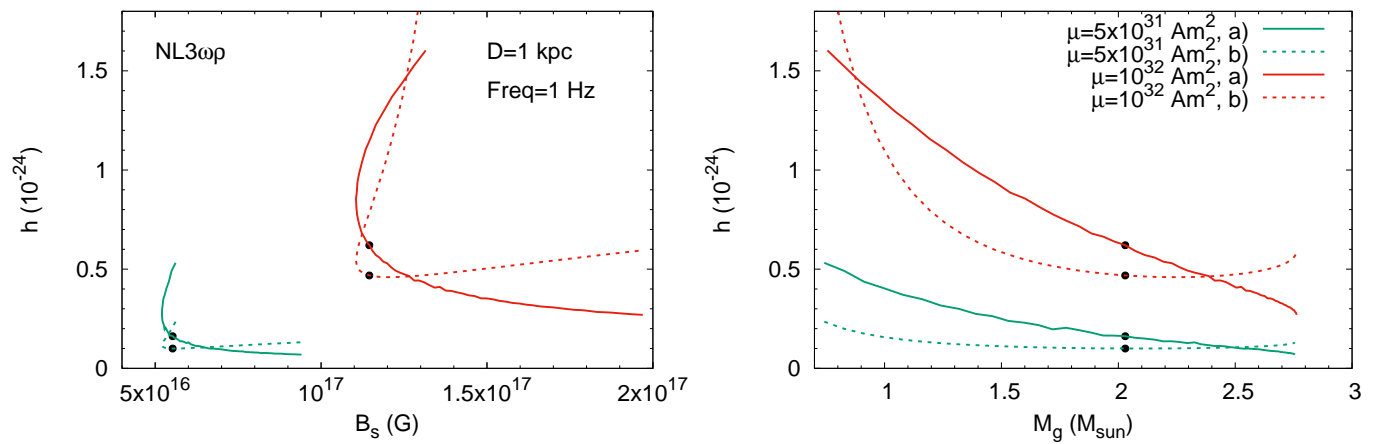} 
 \end{tabular}
  \caption{Gravitational wave amplitude from {Eq.} (\ref{gw-q}) (a) and from Eq. (\ref{gw-B}) (b) as a function of the magnetic field at the surface (pole), $B_s$, (left), and the gravitational mass (right) for the NL3$\omega\rho$ model. We assumed a family of stars located at a distance of 1 kpc and spinning with a frequency of 1Hz. The black dots correspond to the 2-solar-mass star configurations.  } 
\label{fig7}
\end{figure*}

\begin{figure*}[thbp]
 \begin{tabular}{c}
\includegraphics[width=.99\textwidth]{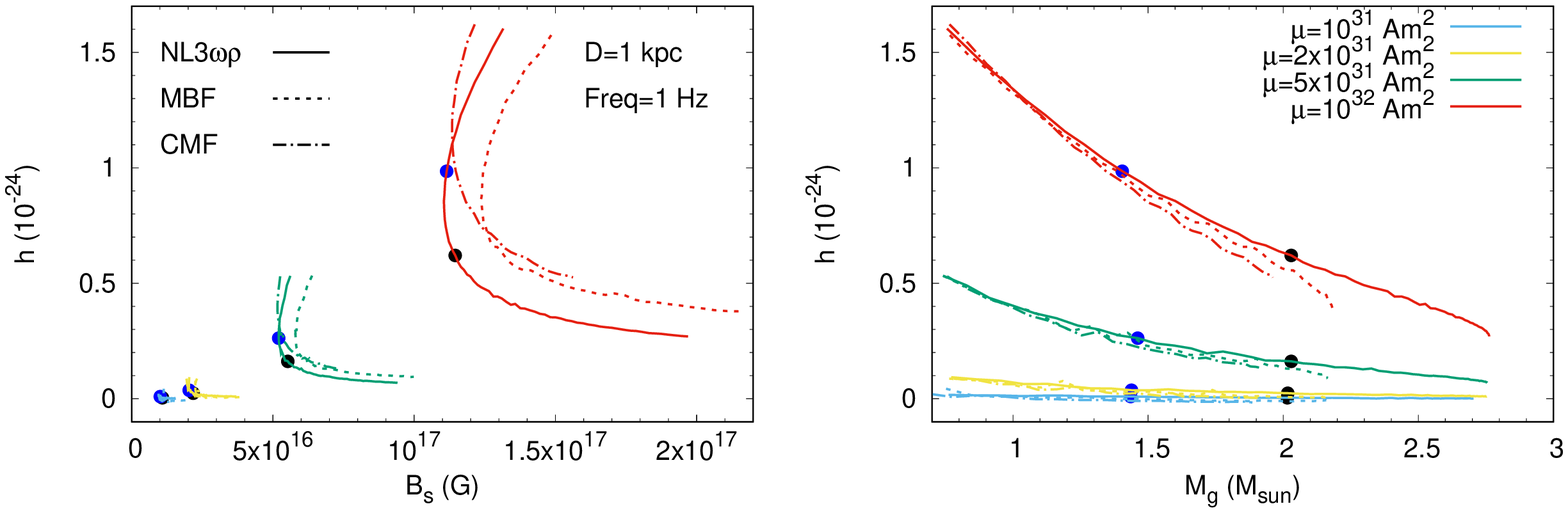} 
 \end{tabular}
  \caption{Gravitational wave amplitude from {Eq.} (\ref{gw-q}), as a function the magnetic field at the surface (pole), $B_s$, (left), and the gravitational mass (right) for the three  models. We once more assumed a family of stars located at a distance of 1 kpc and spinning with a frequency of 1Hz.  The black dots correspond to $\sim$2-solar-mass star configurations, and the blue ones correspond to 1.4 M$_\odot$ stars.}
\label{fig8}
\end{figure*}

In this subsection, Figs. \ref{fig1} to \ref{fig8} show results for several families of stars, using three different EoS: NL3$\omega\rho$, MBF and CMF.  According to  \cite{Haensel2002}, the minimum gravitational mass  of  cold catalysed stars could be as low as $M\sim 0.1$ M$_\odot$, whereas according to \cite{Goussard1998,Strobel1999}, lepton-rich matter as found in proto-neutron stars, is unbound for stars with a mass below $\sim 0.9-1.1$ M$_\odot$. For this reason, in the following, we only show configurations for stars with $M_g > 0.7$ M$_\odot$.

In Fig. \ref{fig1}, we show the gravitational mass of several families of stars with different values of the magnetic dipole moment, $\mu$, as a function of the circumferential radius, which characterises the equator of stars coordinate-independently.  The horizontal bands indicate the mass uncertainties associated with the PSR J0348 $+0432$ \citep{Antoniadis2013} (upper) and PSR J1614$-$2230 \citep{Fonseca16} (lower) masses.
The top panel shows results only for the NL3$\omega\rho$ model. There, one can see that the smaller the mass, the larger the effects of the magnetic field on the stellar structure, as one would expect. A noticeable difference from the non-magnetic case ($\mu=0$) takes place for values of magnetic dipole moment  above $\mu=2\times 10^{31}$ Am$^2$ and only for low-mass stars. This, therefore, implies that, for higher dipole magnetic moments, stars shall have magnetic fields strong enough to make deformation effects non negligible.

 Having this in mind,  results for $\mu=0$, $\mu=5\times 10^{31}$ Am$^2$, and $\mu=10^{32}$ Am$^2$ are displayed for all three  models in the bottom panel. The behavior of the magnetised families of stars calculated with the MBF and CMF models is similar to the one calculated with the  NL3$\omega\rho$ model. Magnetic dipole moments of the order of $5\times 10^{31}$ Am$^2$ and above clearly do have strong effects on the radius of the stars, and this applies to all models.

In Fig. \ref{fig2}, we show the gravitational mass of several families of stars with different values of the magnetic dipole moment, $\mu$, as a function of the surface magnetic field at the pole (top panel), the central magnetic field (middle panel), and the ratio between the central and surface magnetic fields, $B_c/B_s$, (bottom panel), considering the three EoS models. 
Looking at the top panel, one can see that $\mu=5\times 10^{31}$ { Am$^2$}, a value that  gives a significant difference to the $\mu=0$ case in Fig.~\ref{fig1}, corresponds to a surface magnetic field at the pole in the range of $\sim 5\times 10^{16}$ G to $10^{17}$ G for the three models and families of stars considered. These surface fields correspond to a central magnetic field in the range of $\sim 2 \times 10^{17}$ G to $\sim 4\times 10^{17}$ G, as seen from the middle panel.  
In addition, the stellar masses are not significantly affected by the magnetic fields, whereas the radius increases with the $B-$ field, as the stars deform into an oblate shape. The bottom panel shows that, for stars with $M_g \gtrsim 1$ M$_\odot$, we have $B_c\sim 3-5 B_s$. The difference is larger  for low mass stars and weaker fields, i.e. $\mu<2\times 10^{31}$ Am$^2$. This smooth change of magnetic fields inside stars illustrates once more  that ad-hoc exponential parametrizations for the magnetic field \citep{Band97} are unrealistic (as already discussed in \cite{Dexheimer:2016yqu}), as they produce an increase of several orders of magnitude for the magnetic field inside the stars.

The gravitational mass for stellar sequences is plotted  as a function of the central density in Fig. \ref{fig3}.
 The stellar central density is not significantly affected by the magnetic field for the NL3$\omega\rho$: for a 1.8 M$_\odot$ star the density decreases by $\lesssim 0.01$ fm$^{-3}$, while for a 1 M$_\odot$ star, this difference increases to $\lesssim 0.02$ fm$^{-3}$. The CMF stars suffer the larger changes: independently of the stellar mass, as the magnetic field strength changes, the central density decreases by   $\lesssim  0.04$ fm$^{-3}$ when $\mu$ increases from zero to  $\mu= 10^{32}$ Am$^2$. A strong magnetic field pushes the onset of $\Lambda$s and quarks to stars with masses of $\sim 0.08$ M$_\odot$ larger than in  non-magnetised ones, an effect already discussed in  \cite{Rabhi2011}. As a consequence, a nucleonic star may suffer a transition to a hybrid star or a hyperonic star when the magnetic fields decays. The onset of hyperons and quarks  opens new channels for neutrino emission and, therefore, the possibility of a faster cooling process \citep{Yakovlev03,Raduta17,Negreiros18,Grigorian18,Providencia18}. Simultaneously, it also affects the onset of the nucleonic direct Urca process \citep{Fortin16,Providencia18}.   On the other hand, the conversion of a nucleonic or hyperonic star to a hybrid star could be accompanied by the  emission of a reasonable amount of energy, as long as there is a sizable radius change in the involved stellar configurations \citep{Gomes:2018bpw,Dexheimer19}. In this way, the detection of fast cooling or an energetic event coming from a highly magnetised neutron star could be associated with one of these internal degrees of freedom conversions  \citep{Bombaci00,Berezhiani02}.

In Fig. \ref{fig4},  we show the correspondence between the central magnetic field, $B_c$, and the surface magnetic field at the pole, $B_s$, for each value of the magnetic dipole moment considered.  We observe that the family of stars with the lowest $\mu$ considered has a maximum surface magnetic field of $B_s \sim 2\times 10 ^{15}$ G. This is also seen on the top panel of Fig. \ref{fig2}. The sequence of stars with the highest $\mu$ has a corresponding $B_s\sim 2\times 10^{17}$ G and a central magnetic field of the order of  $ 10^{18}$ G. The behaviour of  $B_s$ versus  $B_c$ is not monotonic and, in the bottom panel, the results for $\mu=5\times 10^{31}$ Am$^2$ are shown with more detail for the three EoS considered in this work. The overall behavior is model-independent, although the  quantitative behaviour does depend on the EoS considered.   In the bottom panel, the stars with $M=1.0, 1.4$ and 2 M$_\odot$ have been identified.  The minimum $B_c$ of the curve occurs for a star with a mass of the order of 1.4 M$_\odot$ and corresponds to the star where the mass-radius curve changes curvature.
For stars with a mass  greater or smaller than $\sim1.4$ M$_\odot$, the surface field  increases monotonically with the central magnetic field.  

In Fig. \ref{fig5}, we show the stars' deviation from spherical symmetry, given by the ratio between the radius at the magnetic pole and the radius at the equator, as a function of the gravitational mass. We see that for the lowest magnetic dipole moments considered the deformation is very small and it happens only for the low-mass stars, i.e., with $M<1$ M$_\odot$. For a 1 M$_\odot$ star, the difference between the polar and equatorial radii is below 1$\%$.
Deformation appears for stars with higher magnetic dipole moments. This allows us to conclude that we only have significant deformation  for values of the magnetic dipole moment  $\mu\gtrsim2\times 10^{31}$ Am$^2$, which correspond to a surface magnetic field of $B_s\gtrsim 2 - 4$ ($10^{16}$ G), and $B_c= 1 - 2$ ($10^{17}$ G) (see Fig. \ref{fig2}).

We can also analyse the deformation of stars by looking at their quadrupole moments. Fig. \ref{fig6}  shows the quadrupole moment of the families of stars considered in this study as a function of the gravitational mass. 
As expected, the results for the quadrupole moment are in agreement with the ones shown for the deformation: it becomes non-negligible only for $\mu\gtrsim2\times 10^{31}$ Am$^2$  (see the previous figure).

\begin{figure*}[!htbp]
 \begin{tabular}{c}
\includegraphics[width=.99\textwidth]{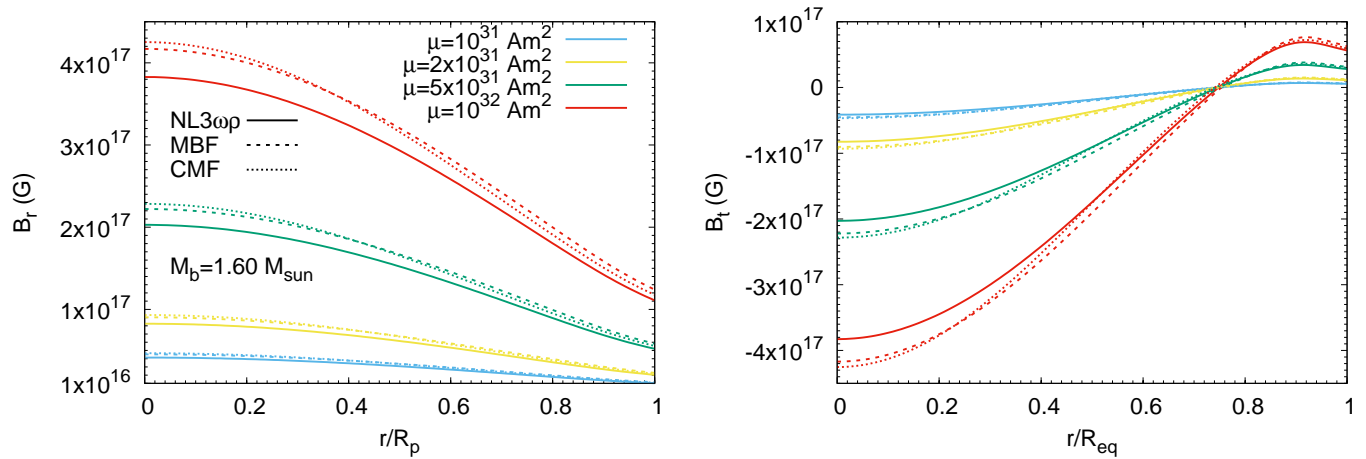} 
 \end{tabular}
  \caption{The radial component of the magnetic field (left) and the tangential component of the magnetic field (right) as a function of the normalised radius for the three  models considering only stars with fixed baryon mass $M_b=1.6$ M$_{\odot}$.  } 
\label{fig9}
\end{figure*}

It has been shown \citep{Bonazzola:1996} that, for a slightly deformed star, the amplitude of gravitational waves emitted is given by
\begin{equation}
h_0=-\dfrac{6G}{c^4}Q\dfrac{\Omega^2}{D} \, ,
\label{gw-q}
\end{equation} 
where $G$ is the gravitational constant, $c$ the speed of light, $Q$ the quadrupole moment, $D$ the distance to the star, and $\Omega$ the rotational velocity of the star.  Another way to estimate the gravitational wave amplitude is from the magnetic field induced deformation \citep{Bonazzola:1996}, where considering an incompressible magnetised fluid, the quadrupole moment can be written as 
\begin{equation}
Q=-\dfrac{\mu_0 \mu^2}{16\pi^2 G \rho R^3} \, ,
\end{equation} 
where $\mu_0$ is the magnetic permeability of free space, so that the ellipticity, $\epsilon_B$, and the moment of inertia, $I$, of the star are given by
\begin{eqnarray}
\epsilon_B&=&\dfrac{15\pi B_s^2 R_{circ}^4}{12\mu_0 G M_g^2} \, , \\
I&=&\dfrac{2}{5}M_g R_{circ}^2 \, ,
\end{eqnarray} 
considering uniform density throughout the star, and approximating  the moment of inertia to the one of a sphere. The gravitational wave amplitude can then be expressed as 
\begin{eqnarray}
h_0=\dfrac{16\pi^2 G \epsilon_B I}{c^4 D P^2} \, ,
\label{gw-B}
\end{eqnarray}
where $P$ is the period of rotation of the star, $P=2\pi/\Omega$.
An estimation of the amplitude of  gravitational waves emitted by these stars is done by setting the distance to 1 kpc (the distances to the pulsars PSR J1614-2230, Vela, and Crab are $\sim 1.2$ kpc , $\sim 0.5$ kpc, and  $\sim 2$ kpc, respectively) and the frequency $f=1/P=1$ Hz.  Results are  shown in the next two figures.

In Fig. \ref{fig7}, we show the gravitational wave (GW) amplitude calculated using the expressions in Eqs. (\ref{gw-q}) and (\ref{gw-B}), for the NL3$\omega\rho$ model. We also mark the 2 M$_\odot$ star configurations. Even though the two curves behave similarly, they do not give the same results and, in fact, the difference is significant: for the  2 M$_\odot$ case,  with $\mu=5\times 10^{31}$ Am$^2$, we obtain $h_0=0.15\times 10^{-24}$ for Eq. (\ref{gw-q}) and $h_0=0.1\times 10^{-24}$ for Eq. (\ref{gw-B}). The difference increases when we consider the $\mu=10^{32}$ Am$^2$ case: $0.65\times 10^{-24}$ for Eq. (\ref{gw-q}) and $0.45\times 10^{-24}$ for Eq. (\ref{gw-B}). The estimation  obtained from Eq. (\ref{gw-B}) is worse for the lower mass stars, which have the larger deformations. This illustrates that using an approximate expression, as in Eq. (\ref{gw-B}), that considers stars with uniform density and a moment of inertia given by the one of a sphere, produces different results and emphasizes the fact that these approaches  should be considered with care when calculating sensitive quantities like GW amplitudes.

In the following, GW amplitudes are estimated from  Eq. (\ref{gw-q}) for the three models we have considered. In Fig. \ref{fig8}, the  gravitational wave amplitude  is shown as a function of the magnetic field at the surface, $B_s$, and as a function of the gravitational mass, for a family of stars located at a distance of 1 kpc and spinning with a frequency of 1 Hz. The black dots correspond to 2 M$_\odot$ configurations and the blue ones correspond to 1.4 M$_\odot$ stars.  

Besides calculating the GW amplitude, we can also estimate the characteristic strain, $S$, the actual quantity measured by the interferometer detectors. It is defined as the product between the GW amplitude and the square root of the integration time \citep{Bonazzola:1996,Moore:2014}: $S=h_0\sqrt{P}$.
If we assume that data was collected during a period of three  years, i.e., $\sqrt{P}\sim 10^4$ s, the  characteristic strain becomes proportional to GW amplitude by a factor of $10^4$.

This means that for a star with $B_s=2\times 10^{16}$ G,  $M_g=2$ M$_{\odot}$, at 1 kpc away and spinning at 1 Hz,  we obtain $h_0=0.15\times 10^{-24}$ and $S=1.5\times 10^{-21}$ Hz$^{-1/2}$. This value could indeed still be detected by detectors like BBO, DECIGO, ALIA. Detectors LIGO and Virgo detect a higher frequency range, $f > 30$ Hz  \citep{gwPlot}.

\begin{figure}[!htbp]
 \begin{tabular}{c}
\includegraphics[width=0.45\textwidth]{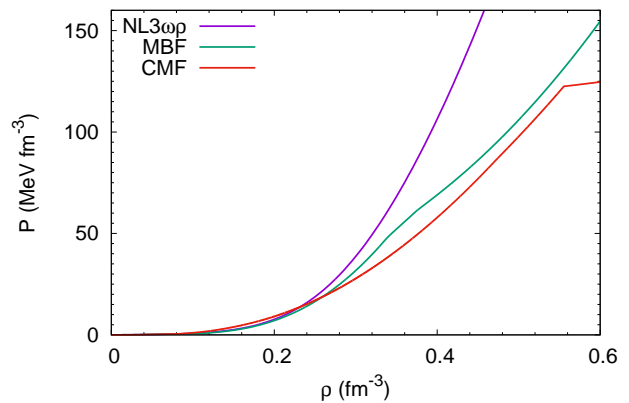} 
 \end{tabular}
  \caption{Equation of state for the three models considered in this work.} 
\label{fig10}

\end{figure}

\subsection{Single-star configurations with $M_b=1.6$ M$_{\odot}$}

\begin{figure*}[!htbp]
 \begin{tabular}{c}
\includegraphics[width=0.99\textwidth]{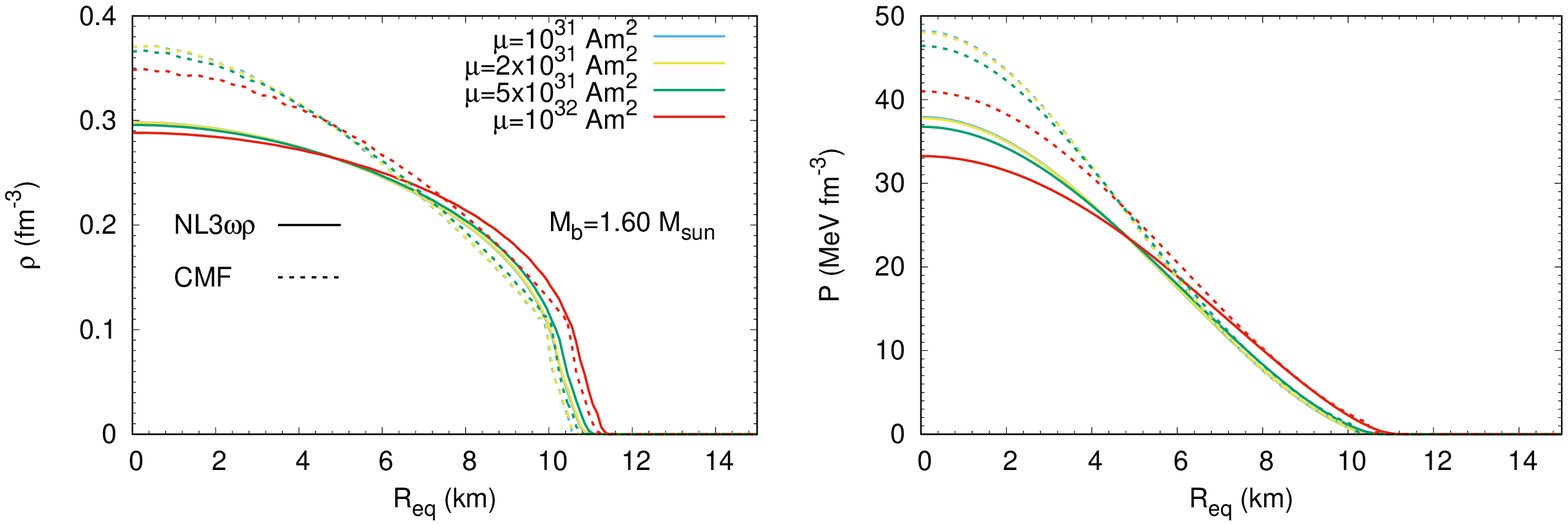} 
 \end{tabular}
  \caption{Profiles of the stars: (left) baryon number density and (right) pressure calculated in the equatorial plane for the three models considered in this work.} 
\label{fig11}
\end{figure*}

In this subsection, we discuss the properties of magnetised stars  with a fixed baryonic mass of 1.6 M$_{\odot}$, with different magnetic dipole moments, $\mu$. Stars with this baryonic mass have a gravitational mass of the order of 1.4 M$_\odot$. The radial component of the magnetic field (left) and the tangential component of the magnetic field (right) are plotted in Fig. \ref{fig9} as a function of the normalised radius for different values of $\mu$. The radial component of $B$, $B_r$, is calculated at $(r,\theta=0,\phi=0)$ (polar) and the tangential (to the equatorial surface) component, $B_t$, is calculated at $(r,\theta=\pi/2,\phi=0)$.  The change of direction of the tangential field occurs for all magnetic momenta at around $R_{eq}/4$ from the surface of the star.  

 As expected, the higher the $\mu$, the higher the magnetic field magnitude. However, it is interesting to see how the intensity of the field depends on the model: NL3$\omega\rho$ has the smallest radial and tangential component absolute values, while the other two models show similar magnitudes. This behavior comes from the fact that the CMF model has the softest EoS and gives rise to larger magnetic field intensities for the same magnetic dipole moment (see Fig. \ref{fig10}, where the three EoS used in this work are represented). From Fig. \ref{fig3}, one can see that stars with $M\sim 1.4$  M$_\odot$ have similar central densities for models CMF and MBF below $\rho=0.4$ fm$^{-3}$. CMF is the softest EoS above saturation and develops the strongest fields in the interior of the star. Below saturation density, however, CMF is closer to NL3$\omega\rho$ and it is the MBF model, with the softest low density EoS,  that develops the strongest surface fields. In the past, it was shown that the compactness of stars is directly related to the strength of central magnetic fields, through the softness of the EoS \citep{Gomes:2017zkc}.  Our results indicate that the relation between the magnetic field in the center of the star and the one on the surface, as discussed in Fig. \ref{fig4}, is defined  by the properties of the  EoS both at the crust and in the core.
 
The effect of magnetic fields on the structure of stars is also clearly shown in Fig. ~\ref{fig11}, where the  profiles of stars with different magnetic field configurations are plotted for models NL3$\omega\rho$ and CMF: the density, and  the pressure are  calculated at the equatorial plane. Effects of the magnetic field are i) a decrease of the density  and pressure in the core of the star and  ii)  an increase of the  pressure at the surface. As a consequence,  matter is pushed outwards and the radius of the star increases. For $\mu=5\times 10^{31}$ Am$^2$, the effect is already seen but it is still small. For $\mu=10^{32}$ Am$^2$, a decrease of the density in the core has as a direct consequence: the suppression of the possible onset of non-nucleonic degrees of freedom,  hyperons or a quark phase, as already discussed in Refs. \citep{Franzon:2015sya,Franzon:2016iai,Gomes:2017zkc}. This explains some results shown in Fig. \ref{fig1} for the CMF model: in the presence of a magnetic field, the onset of $\Lambda$s and quark matter occur in stars with larger masses. 

Let us now consider a star described by the CMF model with strong magnetic field in the interior that has a gravitational (or baryonic) mass above the maximum mass possible for a non-magnetised or weakly magnetised star. A direct consequence of the instability occurring when the magnetic field decays to values that allow for higher central stellar densities, and a possible onset of quark matter, is that the star becomes unstable and decays into a low mass black hole. This transition could be possibly identified by the observation of a gamma-ray burst.

\section{\label{conclusions} Final Remarks}

In the present work, we have calculated the magnetic field strength above which the deformation of the neutron star structure is at least 2\%, which is already non-negligible.
We have considered three representative relativistic mean field models to describe the EoS of neutron stars with diverse compositions (nucleonic, hyperonic and hybrid). The calculations for the structure of stars were performed within the  general relativistic framework implemented in the publicly available Langage Objet pour la RElativité NumériquE (LORENE) library \citep{Bonazzola:1993zz,Bocquet:1995je}.

Our calculation was undertaken by quantifying the deformation due to the magnetic field, in particular, the relation between the polar and equatorial radii, and defining the magnetic dipole moment that causes  a difference below $\sim 1-2 \%$, corresponding to a magnetic field of the order of  $\sim 10^{17}$ G in the center and $\sim 5\times 10^{16}$ G at the surface. It has been shown that, within the adopted formalism,  the magnetic field developed inside the star, and at the surface, depends on the EoS. For a fixed magnetic dipole moment,  stronger magnetic fields are obtained for a softer EoS. Quantitatively, we have found that, for the magnetic field range studied, the relation between these two quantities to be $B_c\sim 3-5\times B_s$, with the relation being dependent on the EoS and the stellar mass.

The determination of the threshold magnetic field for deformed neutron stars was undertaken in two  equivalent approaches. In the polar versus equatorial radii method,  we have identified the magnetic dipole moment that causes a  difference $\sim 1-2 \%$ between both. In the quadrupole moment  method, we use the neutron star quadrupole moment to quantify a  deformation of a similar magnitude. We find that the limiting magnetic field for which the integration of the TOV equations still gives realistic results for the structure of magnetised neutron stars is, in general, model independent.

We report that for stars with masses below $1.5\,\mathrm{M_\odot}$, a dipole magnetic moment of $\sim2\times10^{31}\,\mathrm{Am^2}$, which corresponds to a surface magnetic field of the order of  $\sim 2\times10^{16}$ G and a central magnetic field of  $\sim 8\times 10^{16}$ G, is enough to cause deformation above $1-2\% $ on stars. The same effect is seen for all masses, when stars are described with a dipole magnetic moment of  $\sim5\times10^{31}\,\mathrm{Am^2}$, corresponding to a surface magnetic field of the order of $\sim 5\times10^{16}$ G and a central magnetic field of  $\sim 2-4\times 10^{17}$ G. For the analysis carried out in this work, our results are independent of the model and the composition of stars. 

We have also discussed under which conditions the decay of the magnetic field could give rise to the collapse of the neutron star or the onset of hyperons. Let us refer to the fact that the onset of hyperons inside stars has important effects on the  cooling of the star, not only by affecting the onset density of the nucleonic electron direct Urca process, but also by opening new  direct Urca channels involving hyperons. 

For a matter of completeness, we have investigated the magnetic field distribution inside individual $M_b=1.6$ M$_\odot$ stars, both in equatorial and radial directions, showing that for the case of approximately spherical stars, all the models and compositions present a quantitative and qualitative similar behavior. The models discussed in this work present different baryon density distributions, which is a direct consequence of the degree of stiffness of their EoS. 

Finally, we have compared two methods for determining the gravitational wave amplitude for single stars, considering both the case of approximately spherical stars (slightly deformed) and stars with induced deformation through magnetic effects. We show that, even for 
the lowest magnetic field that generates a relevant deformation, calculations for the gravitational wave amplitude using these two methods never agree. The disagreement between the two methods increases for higher magnetic field amplitudes, as the stars become more deformed. Therefore, we emphasise that such calculations should be carried out cautiously, especially for highly magnetised neutron stars.

\section*{Acknowledgements}
The authors acknowledge support from PHAROS COST Action CA16214 and  by the FCT (Portugal) Projects No. UID/FIS/04564/2016 and POCI-01-0145-FEDER-029912. H.P. was supported by Funda\c c\~ao para a Ci\^encia e Tecnologia (FCT-Portugal) under Project No. SFRH/BPD/95566/2013, and VD by the National Science Foundation under grant PHY-1748621.

\end{document}